\documentclass[sigconf]{acmart}
\AtBeginDocument{%
  }

\copyrightyear{2026}
\acmYear{2026}
\setcopyright{cc}
\setcctype{by}
\acmConference[UbiComp Companion '26]{Companion of the 2026 ACM International Joint Conference on Pervasive and Ubiquitous Computing}{October 11--15, 2026}{Shanghai, China}
\acmBooktitle{Companion of the 2026 ACM International Joint Conference on Pervasive and Ubiquitous Computing (UbiComp Companion '26), October 11--15, 2026, Shanghai, China}
\acmDOI{10.1145/3798063.3837248}
\acmISBN{979-8-4007-2533-3/2026/10}

\usepackage{booktabs}
\usepackage{tabularx}
\usepackage{array}
\usepackage{multirow}
\usepackage{makecell}
\usepackage{ragged2e}
\usepackage{graphicx} 
\usepackage{subcaption} 
\usepackage{xcolor}
\usepackage{tabularray}
\usepackage{float}
\usepackage{balance}

\begin{document}

\title[Health Ikigai]{Health Ikigai: Semantic Conversational Interfaces for Wearable Health Data to Support Behavioral Change for Senior Indian}


\author{Md Haseen Akhtar}
\authornote{Md Haseen Akhtar and Hansoo Lee contributed equally to this work as joint lead authors.}
\affiliation{%
  \institution{Imperial College London}
  \country{United Kingdom}
}
\email{m.akhtar1@imperial.ac.uk}
\affiliation{%
  \institution{IIT Hyderabad}
  \country{India}
}
\email{haseen@des.iith.ac.in}

\author{Hansoo Lee}
\authornotemark[1]
\authornote{Corresponding author.}
\affiliation{%
  \institution{Imperial College London}
  \country{United Kingdom}
}
\email{h.lee1@imperial.ac.uk}
\affiliation{%
  \institution{KIST}
  \country{Republic of Korea}
}
\email{hansoolee@kist.re.kr}

\author{Rafael A. Calvo}
\affiliation{%
  \institution{Imperial College London}
  \country{United Kingdom}
}
\email{r.calvo@imperial.ac.uk}

\renewcommand{\shortauthors}{Md Haseen Akhtar, Hansoo Lee, and Rafael A. Calvo}
\begin{abstract}
In ubiquitous health contexts, wearable feedback is often encountered in brief, interrupted moments of everyday life. However, many wearable and mobile health systems still present personal health data through numerical dashboards, leaving users and health coaches to interpret trends, thresholds, and deviations on their own. This interpretive burden may be especially relevant for adults aged 50--65 in India, where digital literacy, health literacy, language, and daily routines can shape how health technologies are understood and used. We present Health Ikigai, a design-oriented proof-of-concept framework that reframes wearable activity feedback from dashboard-based numerical interpretation to semantic conversational feedback. The prototype transforms replayed wearable-derived activity data into baseline-aware triggers, plain-language feedback, limited-turn conversational prompts, and coach-facing review summaries. A preliminary formative evaluation with ten adults in India suggests that participants perceived the interaction as understandable, lightweight, supportive, and actionable. We discuss design implications for low-burden wearable health interfaces that preserve coach oversight while reducing users' interpretive work.
\end{abstract}





\maketitle

\section{Introduction}
Wearable devices increasingly collect continuous health-related signals such as physical activity, step counts, heart rate, sleep duration, inactivity, and other behavioral or physiological indicators~\cite{piwek2016rise, huhn2022impact}. However, access to continuous personal health data does not automatically lead to reflection or behavior change. Prior work in personal informatics and wearable/mobile based self-tracking has shown that users must interpret what their data means, decide whether a pattern is meaningful, and connect that interpretation to feasible everyday action~\cite{li2010stage, li2011understanding, epstein2015lived, kersten2017personal, lee2026sage, lee2025leveraging}. Wearable health feedback is therefore not only a sensing problem, but also a sensemaking and interaction design problem.

Many wearable and mobile health systems still present personal health data through dashboard-first interfaces that foreground numerical indicators, trend charts, thresholds, and summary views~\cite{faisal2013making, turchioe2019systematic}. While such interfaces can support analysis, they often leave users to determine whether a deviation is important, what may have caused it, and what action should follow~\cite{epstein2015lived, kersten2017personal, cocskun2023data}. We use the term interpretive burden to describe this cognitive work of transforming wearable metrics into everyday meaning and feasible action. This burden can increase when users must interpret baselines, trends, and deviations during brief, interrupted moments of daily life rather than sustained analytic review~\cite{cocskun2023data, jones2018dealing}. Prior work also suggests that users need scaffolding to connect personal health data to self-management decisions~\cite{mamykina2008mahi}.

This challenge is especially relevant for adults aged 50--65 in India, who may face elevated risks related to diabetes, hypertension, obesity, and physical inactivity~\cite{anjana2023metabolic, pengpid2022prevalence}. At the same time, differences in digital literacy, health literacy, language, and everyday routines can shape how health technologies are understood and used~\cite{joshi2023health, rasekaba2022exploring}. For people who check wearable feedback while commuting, working, caregiving, or resting, dense numerical dashboards may be less appropriate than simple feedback based on personal baselines and small feasible actions.

\begin{figure}
    \centering
    \includegraphics[width=\linewidth]{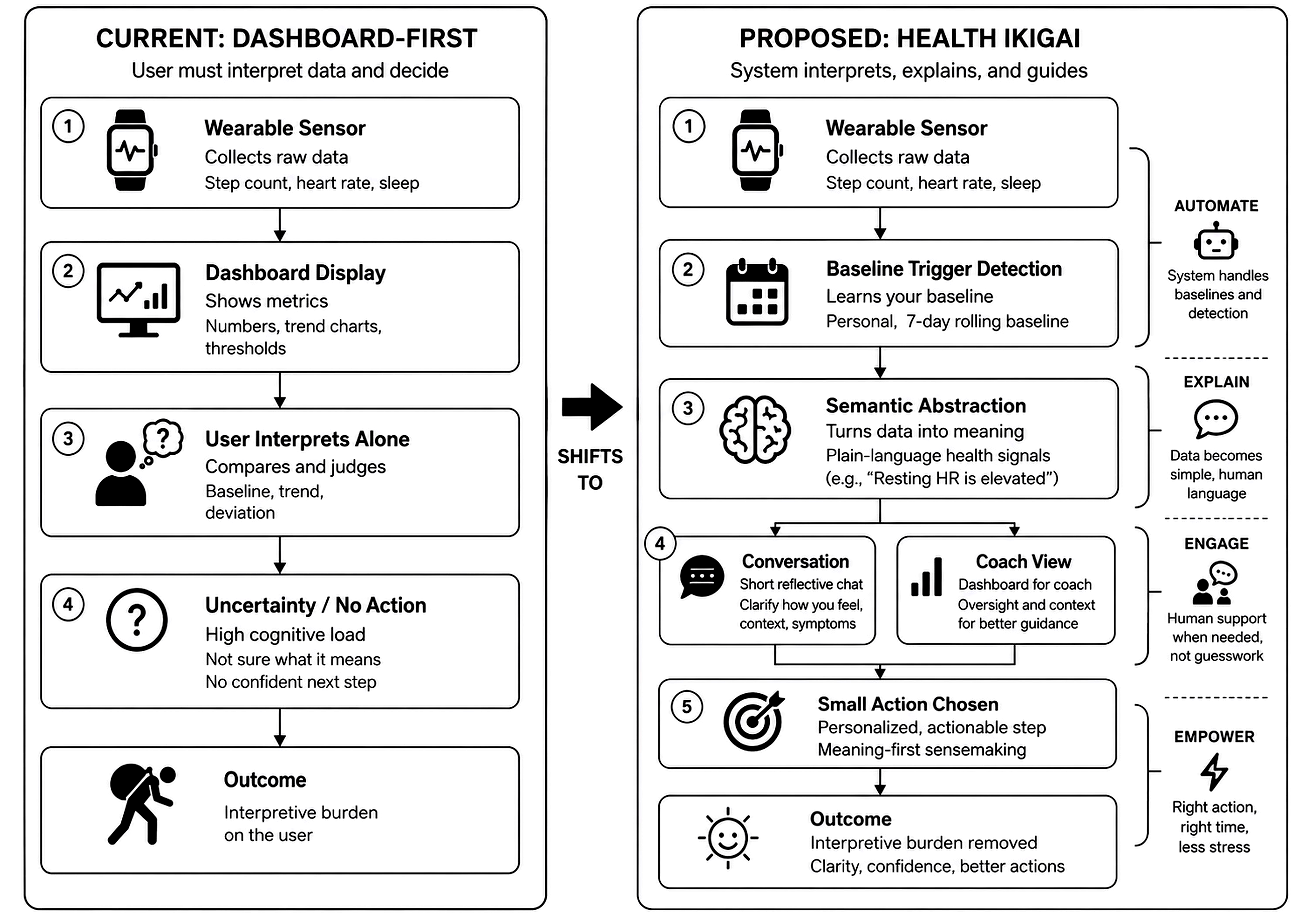}
    \caption{Health Ikigai reframes wearable feedback from dashboard-first numerical interpretation (left) to meaning-first conversational sensemaking with coach-facing oversight (right)}
    \label{fig:health-ikigai}
\end{figure}

In response, we present Health Ikigai, a design-oriented proof-of-concept framework that reframes wearable feedback from dashboard-first numerical interpretation to meaning-first conversational sensemaking. As shown in Figure~\ref{fig:health-ikigai}, the system shifts the primary user experience away from dashboards that require users to interpret trends, thresholds, and deviations on their own, and toward brief semantic conversations that explain what a change may mean and support small feasible actions. The dashboard is not removed; it is repositioned as a coach-facing supervisory layer where health coaches can review baseline settings, trigger states, conversation context, and underlying step records when needed. Drawing on prior work on natural-language explanations of statistical patterns~\cite{bentley2013health} and clinician-centered wearable dashboards~\cite{sadhu2023designing}, Health Ikigai translates Apple Health-derived step-count records into baseline-aware triggers, semantic health signals, short conversational prompts, and coach-facing summaries. This paper addresses two research questions. RQ1: How can wearable health data be reframed from dashboard-first numerical presentation to meaning-first conversational feedback for middle-aged and older adults in India? RQ2: What conversational-semantic design principles can reduce the cognitive complexity of wearable health data for this population?

Health Ikigai should be understood as a proof-of-concept rather than a validated health intervention. The current prototype uses replayed Apple Health step-count records to demonstrate an interaction pipeline from activity data to trigger detection, semantic abstraction, limited-turn conversation, and coach-facing review. A preliminary formative evaluation with ten participants aged 50–65 in India provides early evidence that this interaction style was perceived as clear, supportive, and actionable. We make three contributions: (1) framing dashboard-first wearable feedback as interpretive burden in preventive activity support; (2) presenting a layered proof-of-concept system connecting baseline-aware trigger detection, semantic abstraction, limited-turn conversation, and coach-facing oversight; and (3) reporting preliminary formative findings on plain-language semantic feedback and bounded action-oriented conversation.

\section{Design Rationale and System Overview}
\label{sec:designrationale}

Health Ikigai was developed through a user-centered design process comprising persona creation, journey mapping, and feature ideation. We defined our target users as adults aged 50--65 who are managing weight and daily health routines in relation to age-related conditions or lifestyle transitions. These users may own wearable devices but find complex numerical dashboards difficult to interpret during routine daily activity, consistent with prior work on technology acceptance among older adults~\cite{chung2023community, mitzner2010older}.

\begin{figure}
    \centering
    \includegraphics[width=\linewidth]{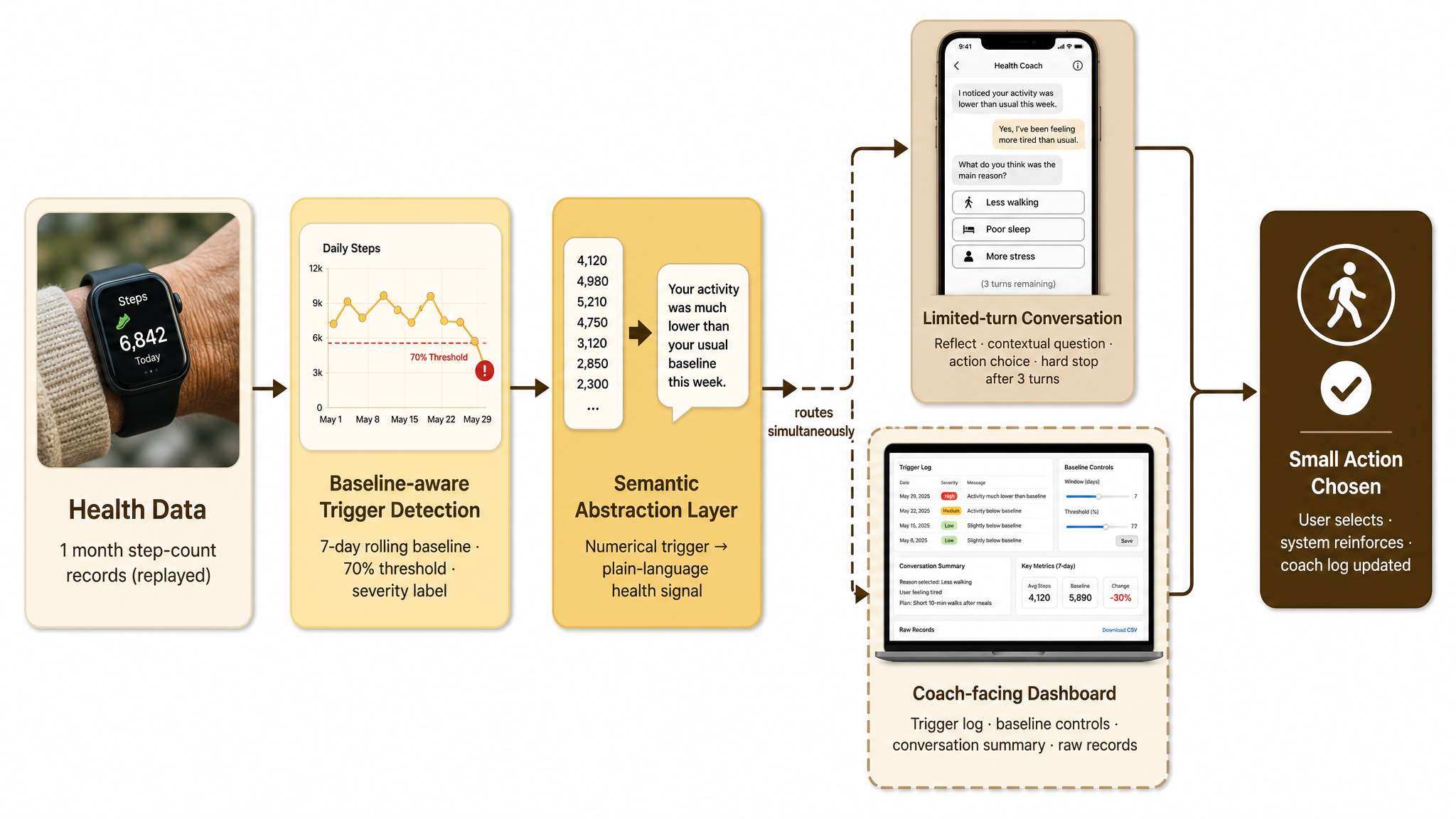}
    \caption{Health Ikigai system pipeline transforming wearable data via trigger detection, semantic abstraction, and reflective conversation. The dashed arc shows semantic signals feeding both user and coach interfaces}
    \label{fig:end-to-end-pipeline}
\end{figure}

We developed a persona around a small businessman in India aged 50--65, who wants to improve health through regular walking but finds wearable dashboards confusing, has limited time for routine data review, and prefers simple meaningful feedback revealing design priorities around autonomy, dignity, non-judgmental language, and advice that fits real routines. Journey mapping across daily routines, informed by prior work on routine-based mobile sensing~\cite{lee2025leveraging}, revealed that the system should support reflection and offer small achievable actions rather than issue instructions in response to baseline deviations, with feedback that fits naturally into daily rhythms. Observations and insights were converged into six design requirements: feedback should be meaning-first, metric-second; interpretation should be grounded in personal baselines rather than population norms; prompts should be non-judgmental and respectful; interactions should be lightweight and easy to parse; the dashboard should support coach oversight without burdening users; and the system should preserve the user's sense of control and independence. These themes are consistent with literature showing that technology acceptance and sustained use among older adults are shaped by perceived simplicity, social meaning, and alignment with personal values and routines~\cite{mitzner2010older, fowe2022understanding}. Ideation produced proposals across three areas: coach-facing dashboard support, motivational feedback for users, and interaction design emphasising plain language over medicalized phrasing and emotionally safe communication that avoids blame~\cite{lee2026front}.

Health Ikigai presents meaning-first interaction to users while preserving reviewability for coaches through dashboard inspection and details-on-demand (see Figure~\ref{fig:end-to-end-pipeline}). Health Ikigai uses the semantic layer to drive a short conversational behavior-change support flow. In the current proof-of-concept, this flow follows a reflect-and-evoke principle. Once a trigger has been detected, the system begins by reflecting the observed state back to the user in simple language. For example, instead of saying ``\emph{your steps are below baseline by 180},'' the system may say ``\emph{This week feels a little lighter than usual on movement}.'' It then follows with a reflective question such as ``\emph{What has been different today?}'' If the user replies with contextual information such as ``\emph{I have been busy},'' the system shifts into an evocation mode. Rather than prescribing a single response, it offers two or three feasible alternatives and asks which feels most realistic. For example, it may suggest a short walk after lunch, a few brief movement breaks between tasks, or another low-burden option compatible with the user's routine. If the user selects one of these options, the system reinforces the choice and intentionally stops encouraging manageable action.

\newcolumntype{Y}{>{\RaggedRight\arraybackslash}X}

\definecolor{normalgreen}{RGB}{95,160,80}
\definecolor{loworange}{RGB}{230,140,35}
\definecolor{verylowred}{RGB}{170,0,0}
\definecolor{elevatedblue}{RGB}{0,0,255}

\begin{table}
\centering
\scriptsize

\begingroup
\renewcommand{\tabularxcolumn}[1]{m{#1}}

\begin{tabularx}{\columnwidth}{@{}YYYY@{}}
\toprule
\textbf{Numerical Condition} &
\textbf{Semantic Abstraction} &
\textbf{User-Facing Output} &
\textbf{Coach-facing Explanation} \\
\midrule

\textbf{\textcolor{normalgreen}{Normal}}\par
Steps $\geq$ 70\% of 7-day baseline
&
Typical Activity Day
&
\textit{``You have been moving regularly lately, keep it up.''}
&
No intervention needed. Selected day within expected range.
\\
\midrule

\textbf{\textcolor{loworange}{Low activity}}\par
Steps 50--69\% of baseline
&
Slightly less activity than usual
&
\textit{``This week feels a little lighter than usual with movement. What has been different today?''}
&
Mild deviation from baseline; reflective check-in triggered. Users are shown two or three small action options.
\\
\midrule

\textbf{\textcolor{verylowred}{Very low}}\par
Steps $<$ 50\% of baseline
&
Notably reduced activity
&
\textit{``You seem to have moved quite a bit less than usual lately. Is everything okay?''}
&
Meaningful drop detected; coach flag raised. The system prompts reflection and feasible alternatives; coaches may follow up directly.
\\
\midrule

\textbf{\textcolor{elevatedblue}{Elevated}}\par
Steps $\geq$ 130\% of baseline
&
More active than usual
&
\textit{``You have been moving quite a bit more than usual, great effort! Remember to rest if you need it.''}
&
Above-average activity noted; positive reinforcement delivered. Reminder to balance exertion with recovery.
\\
\midrule

Multi-day trend\par
3+ consecutive below-threshold days
&
Sustained reduced movement pattern
&
\textit{``Over the past few days your movement has been quieter than usual. Would a small routine adjustment help?''}
&
Persistent low-activity pattern detected across multiple days; coach escalation flag raised for optional human follow-up.
\\

\bottomrule
\end{tabularx}

\endgroup

\caption{Numerical-to-semantic mapping linking trigger conditions to semantic abstractions, user-facing messages, and coach interpretations}
\label{tab:numerical-semantic-mapping}
\end{table}

In the current proof-of-concept, the prototype uses approximately one month of step-count records generated by smart watches and stored in the health app, collected via a custom logger. These records are replayed sequentially rather than streamed in real time, allowing the prototype to demonstrate the full interaction pipeline, from activity data to trigger to semantic message to conversational exchange without requiring a live wearable connection. The following subsections describe each layer of the pipeline in turn. The architecture was developed around behavior-change-oriented interaction design, with emphasis on semantic interpretation, feasible action suggestions, and lightweight coaching support~\cite{sadhu2023designing, cocskun2023data}. The main contribution of Health ikigai is the interplay and communication from numerical layer to the semantic layer.

The semantic abstraction layer translates numerical trigger conditions into plain-language health signals before user-facing interaction begins. For example, a date with above-baseline activity produces a positive acknowledgment with no intervention; a below-baseline date generates a low-activity trigger, a dashboard alert, and a conversational prompt. The user can respond in natural language, receiving two or three feasible alternatives, after which the system reinforces the chosen option and encourages real-world action rather than prolonged chat~\cite{cocskun2023data, ma2025making}. The translation follows a consistent mapping logic: the severity and direction of a trigger determine a semantic phrase category, which in turn shapes both the conversational opening and the coach-facing dashboard state. Table~\ref{tab:numerical-semantic-mapping} summarizes the full mapping.

\begin{figure}
    \centering
    \includegraphics[width=\linewidth]{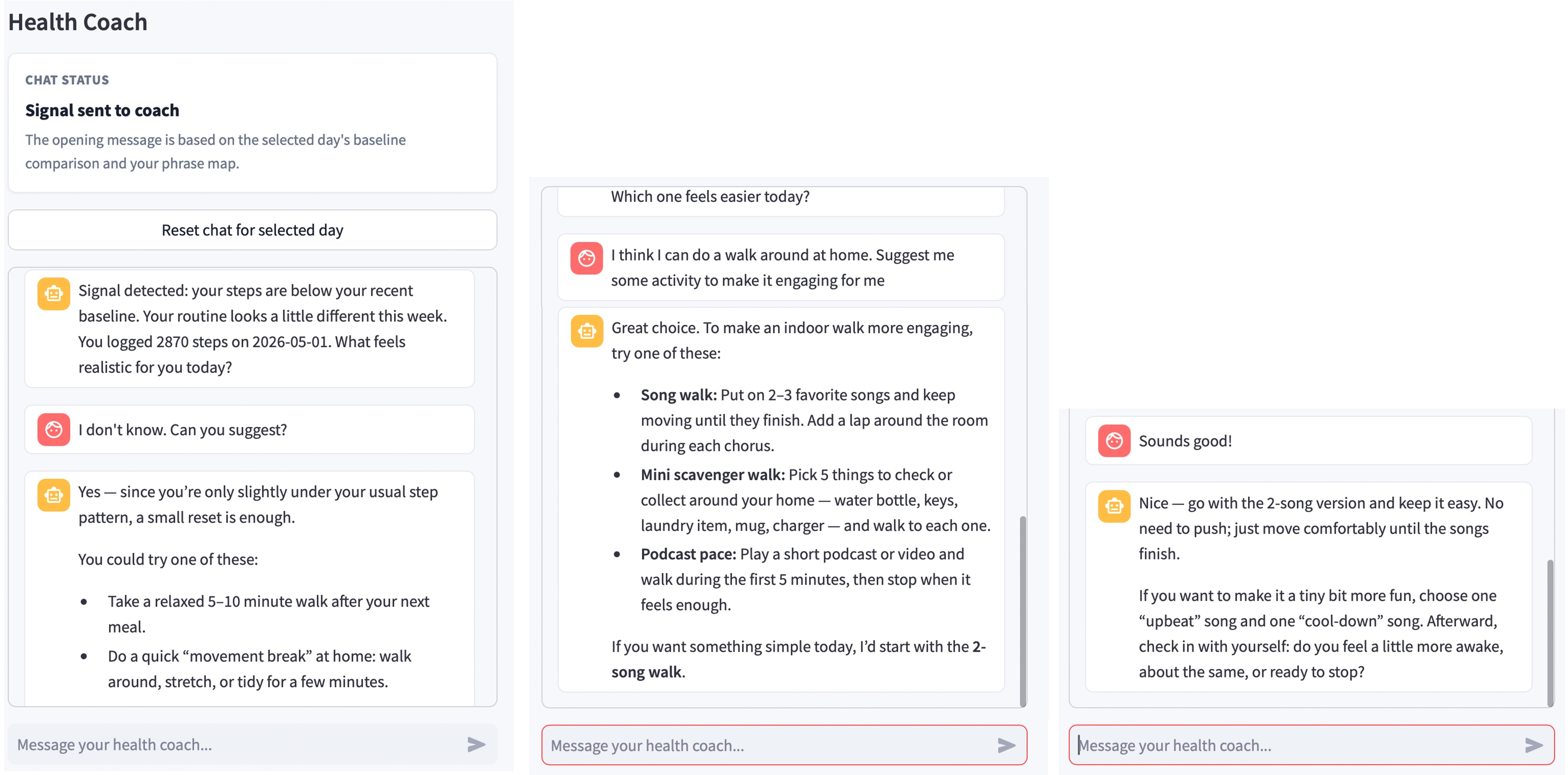}
    \caption{\textit{Example trigger-to-conversation flow}: A below-baseline activity day is translated into a structured low-activity trigger, a semantic phrase, and a short reflective coaching exchange that supports a small user-chosen action.}
    \label{fig:trigger-to-conversation}
\end{figure}

\begin{figure}
    \centering
    \includegraphics[width=\linewidth]{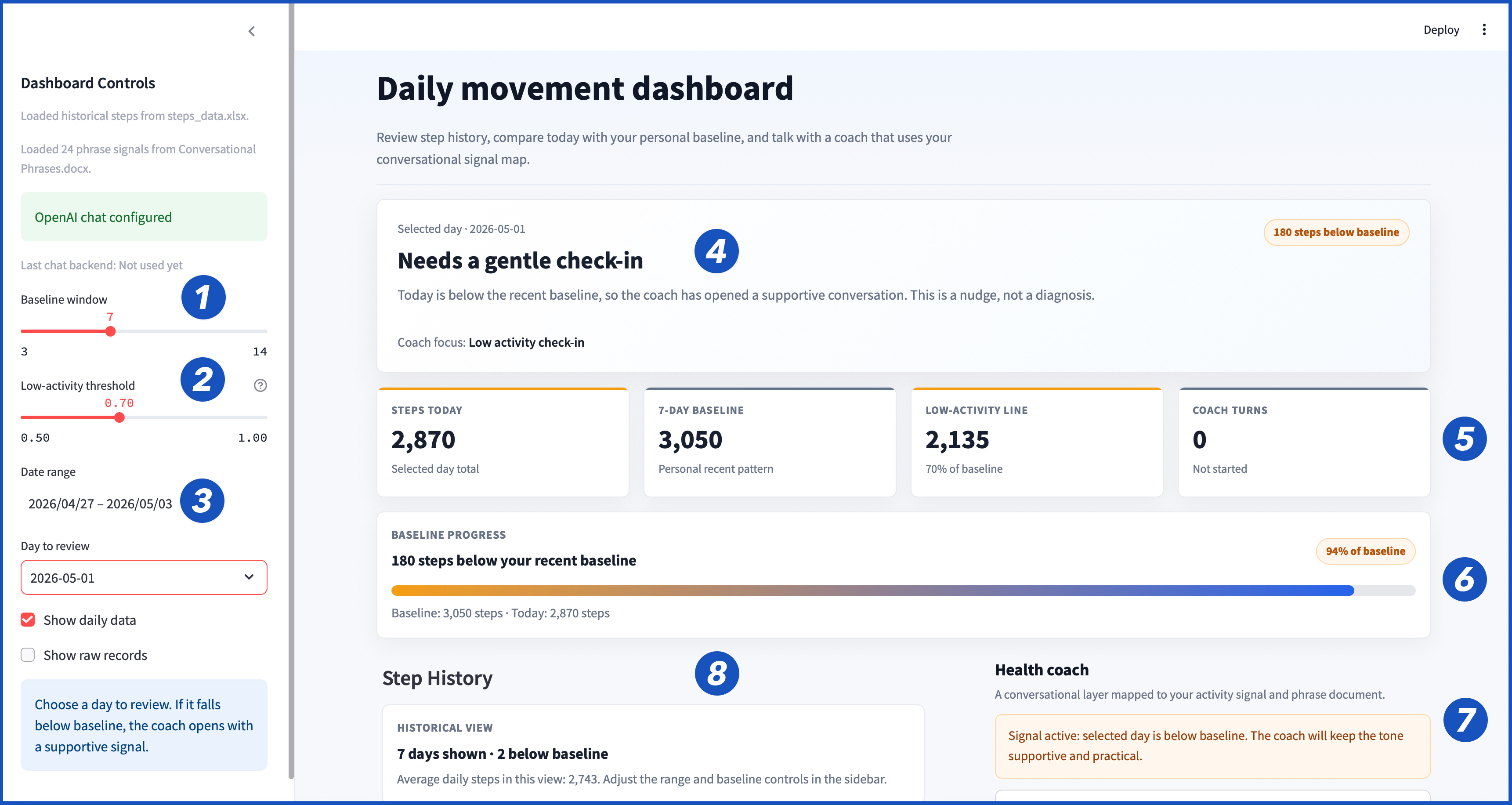}
    \caption{Coach dashboard components: (1) baseline/history controls; (2) threshold slider (e.g., 70\% for low activity); (3) analysis date range; (4) semantic trigger state; (5) metric cards (steps, baseline, activity type, turns); (6) progress bar; (7) conversation context (trigger, user input, chosen action); and (8) 7-day step history}
    \label{fig:coach-dashboard-callout}
\end{figure}

The design philosophy is that the system should help users notice a pattern and choose a feasible action, not draw them into extended chat~\cite{lee2026front}. Closing the conversation after a small set of turns was a deliberate design choice and, as reported in Section~\ref{sec3}, was one of the most positively received features of the prototype. Figure~\ref{fig:trigger-to-conversation} illustrates this pipeline with a concrete example. The coach-facing dashboard allows health coaches and review stakeholders to inspect how a trigger was produced and how the conversational response unfolded, without requiring them to review raw records first (See Figure~\ref{fig:coach-dashboard-callout})~\cite{wang2022usability, sadhu2023designing}.

\section{Study Design}
\label{sec3}
We conducted a preliminary formative evaluation to assess whether the prototype's semantic and conversational design was understandable, actionable, and perceived as supportive by the target population. Ten participants were recruited from India, all in the 50--65 age range (eight men, two women). Several participants reported managing everyday health conditions including diabetes and high blood pressure, reflecting characteristics relevant to the broader target population for cardiometabolic preventive support. Recruitment inclusion criteria required that participants had prior difficulty interpreting numerical health information directly, as assessed through a short screening question. Informed consent was obtained from all participants before they interacted with the prototype and completed the survey.

Participants interacted with the prototype individually through a facilitated session in which a low-activity scenario was replayed and the conversational interface was demonstrated. Each participant engaged with the prototype directly, responding to the system's prompts as they would in everyday use. Following the interaction, participants completed a ten-item questionnaire and provided open-ended verbal feedback. Sessions were conducted in a language comfortable to each participant. All participants provided informed consent prior to participation. The study focused on assessing whether participants could understand the semantic feedback, perceive the suggested actions as feasible, and experience the interaction as supportive, lightweight, and easy to follow. 

\section{Preliminary Results}
\begin{figure} 
    \centering 
    \begin{subfigure}{\linewidth} 
        \centering 
        \includegraphics[width=\linewidth]{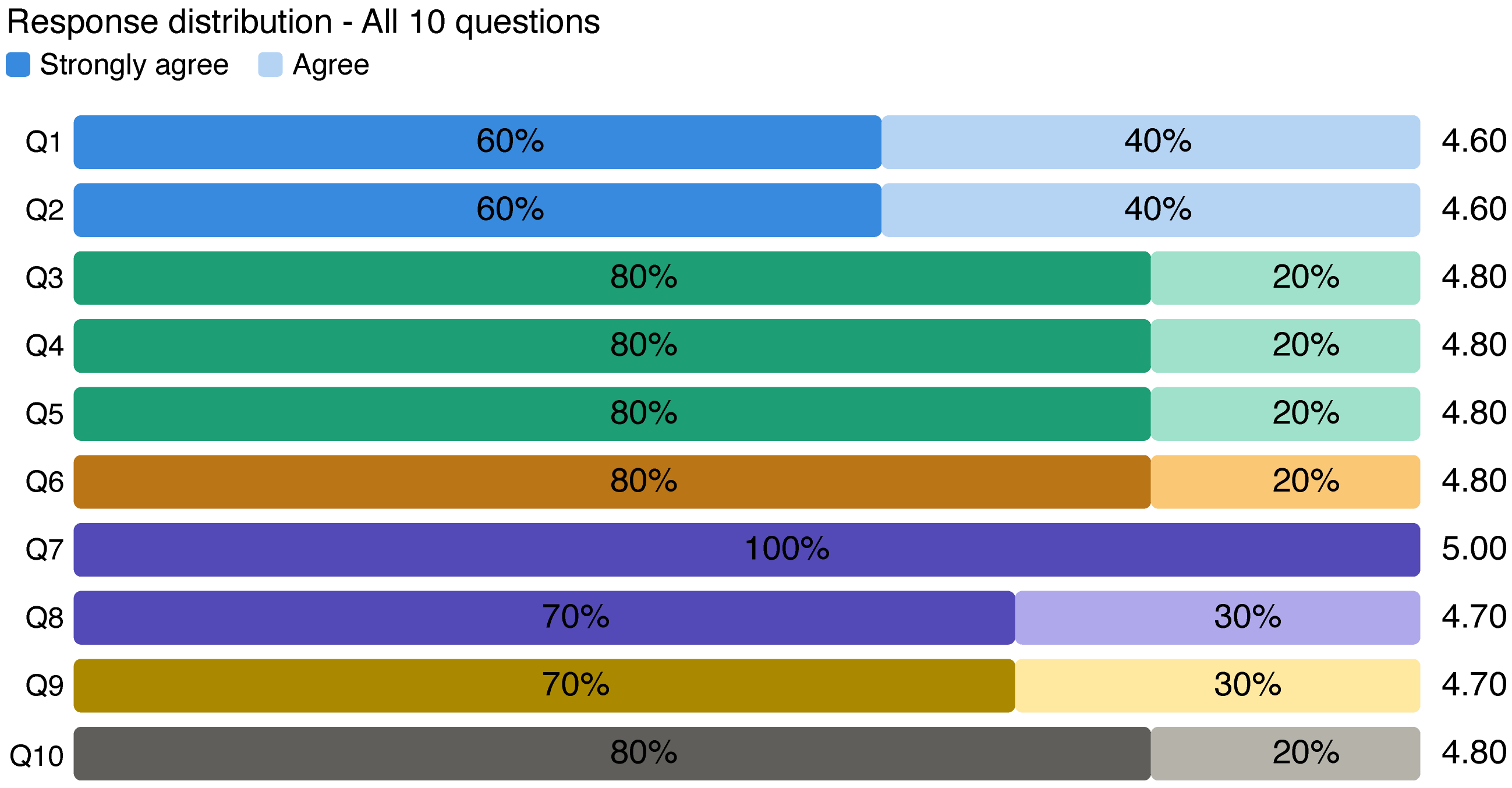} 
        \caption{Response distribution across all ten questions.} \label{fig:survey-results-all} 
    \end{subfigure} 
    
    \begin{subfigure}{\linewidth} 
        \centering 
        \includegraphics[width=\linewidth]{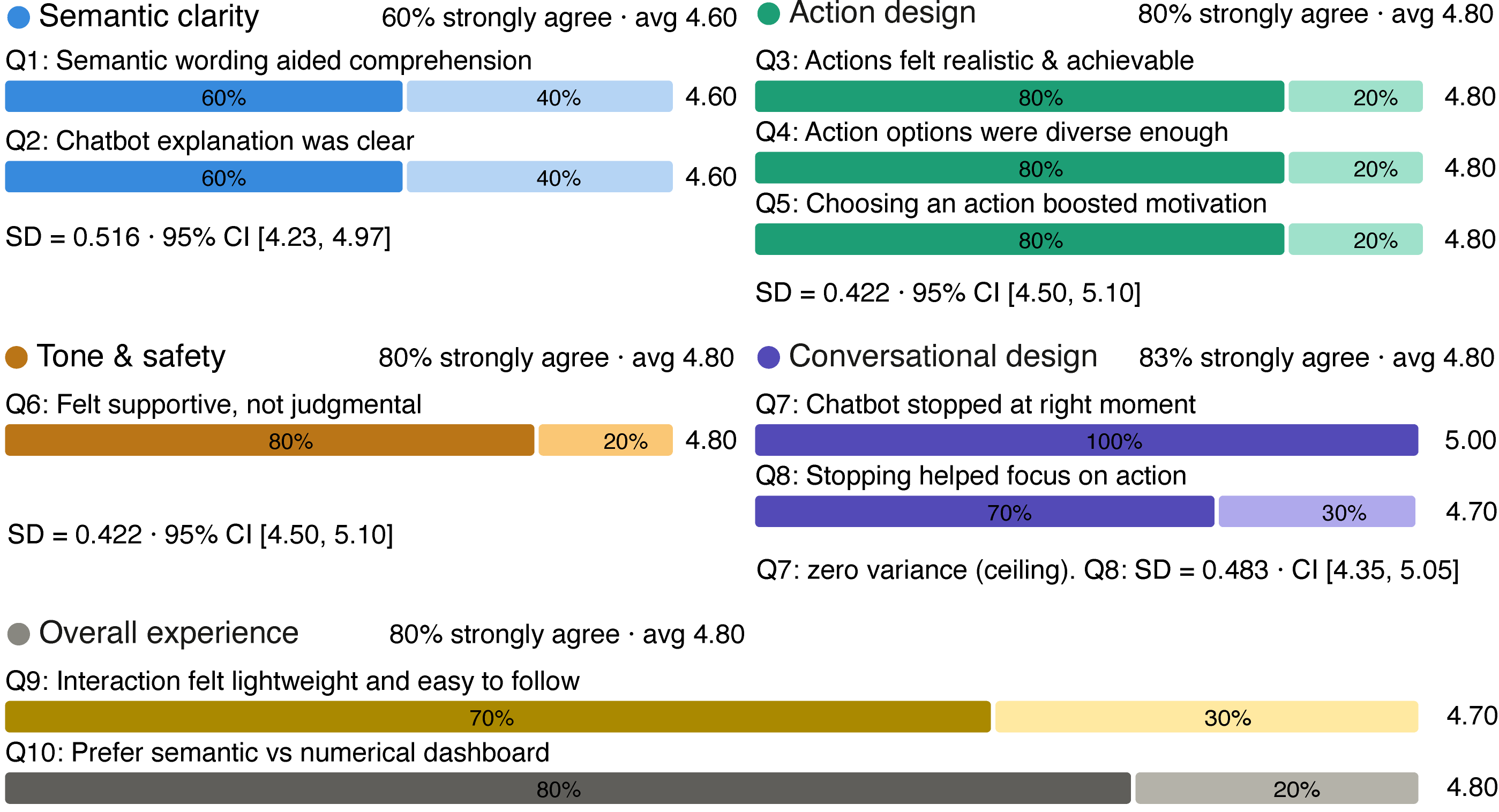} 
        \caption{Results grouped by design theme.} 
        \label{fig:survey-results-theme} 
    \end{subfigure} 
    \caption{Survey results for 10 participants. (a) Overall response distribution, showing a mean score of 4.76 and a strong concentration of positive responses across clarity, actionability, tone, and limited-turn conversation design. Q7 achieved a perfect score of 5.0, while Q8 scored 4.70. (b) Results grouped by theme: Semantic Clarity (Q1--Q2), Action Design (Q3--Q5), Tone and Safety (Q6), Conversational Design (Q7--Q8), and Overall Experience (Q9--Q10).} 
    \label{fig:survey-results} 
\end{figure}

Responses were coded on a five-point scale (1 = Strongly Disagree, 5 = Strongly Agree), with an overall mean of 4.76 across all items. Semantic clarity (Q1--Q2) returned a mean of 4.60 ($SD = 0.516$, 95\% CI [4.23, 4.97]), with participants reporting that plain-language wording helped them understand health states without interpreting raw numbers or trend charts. Action design and tone (Q3--Q6) scored a mean of 4.80 ($SD = 0.422$, 95\% CI [4.50, 5.10]), with participants valuing multiple distinct action options as more motivating and respectful than a single prescribed directive. Conversational design items showed the strongest endorsement: Q7 (``chatbot stopped at an appropriate point'') achieved a perfect mean of 5.00 with zero variance, and Q8 (``stopping helped focus on action'') returned 4.70 ($SD = 0.483$, 95\% CI [4.35, 5.05]), suggesting the turn limit was perceived as an intentional and effective design feature rather than a restriction. Overall experience items (Q9--Q10) returned means of 4.70 and 4.80 respectively. The preference item for semantic conversational feedback over numerical dashboard presentation was also highly rated within the demonstrated prototype context, scoring 4.80 (see Figure~\ref{fig:survey-results}).

\section{Discussion and Conclusion}
Health Ikigai reframes wearable health feedback as a sensemaking problem rather than only a visualization problem. By separating backend numerical computation from frontend communication, the prototype translates step-count deviations into plain-language feedback while preserving numerical transparency for coach review. In response to RQ1, this design suggests one possible way to shift interpretive work away from users without removing the dashboard entirely. Instead, the dashboard is repositioned as a coach-facing supervisory layer where coaches can inspect baseline settings, trigger states, conversation summaries, and underlying step records when needed. The preliminary formative results suggest that participants perceived the semantic conversational interaction as clear, supportive, and actionable, particularly when the conversation stopped after helping them choose a feasible next action. Rather than treating the chatbot as an open-ended companion, Health Ikigai uses bounded conversation as a lightweight bridge from sensed activity change to reflection and small action planning.

In response to RQ2, the design highlights conversational-semantic requirements for low-burden wearable feedback in contexts such as India, where language, health literacy, digital confidence, daily routines, and cardiometabolic risk may shape how personal health data is understood and acted upon. The current work remains an early-stage proof-of-concept. The prototype uses replayed wearable-derived step-count records rather than live wearable streams, focuses on a single activity metric, and was evaluated with a small formative sample without a comparison condition or health outcome measures. Future work should evaluate Health Ikigai with larger and more diverse groups, live multimodal wearable data, and controlled comparisons with conventional user-facing dashboards.

\begin{acks}
This project was partially supported by the Sejong Science Fellowship [RS-2025-00559234], funded by the National Research Foundation (NRF) of Korea.
\end{acks}
\balance
\bibliographystyle{ACM-Reference-Format}
\bibliography{sample-base}


\end{document}